\documentclass[12pt]{article}
\usepackage{amssymb}
\usepackage{epsfig}
\usepackage{float}
\usepackage{graphicx}

\begin{document}
\begin{center}
{\bf On the theory of neutrino mixing and oscillations}\footnote{Lectures given at IVth International Pontecorvo Neutrino Physics School (26.09-06.10, 2010, Alushta, Crimea, Ukraine).}
\end{center}
\begin{center}
S. M. Bilenky
\end{center}

\begin{center}
{\em  Joint Institute for Nuclear Research, Dubna, R-141980,
Russia.\\}
\end{center}
\begin{center}
{\em Physik- Department E15, Technische Universit\"at M\"unchen, 

D-85748, Garching, Germany}
\end{center}
\begin{abstract}
 A brief review of the status of neutrino oscillations is given. The phenomenology of  neutrino mixing and the standard seesaw mechanism of neutrino mass generation is discussed. Different approaches to neutrino oscillations are considered and compared. The role of the Heisenberg space-momentum uncertainty relation and the Mandelstam-Tamm time-energy uncertainty relation in neutrino oscillations is discussed in some detail.
\end{abstract}

\section{Introduction. Present status of neutrino oscillations}

The observation of neutrino oscillations in the solar, atmospheric, reactor and accelerator neutrino
experiments \cite{Solar,SK,Kamland,K2K,Minos} is one of the most important recent discoveries in  particle physics. Small neutrino masses and peculiar neutrino mixing are commonly considered as a signature of  new physics  beyond the Standard Model.

Existing neutrino oscillation data (with the exception of the LSND \cite{LSND} and recent MiniBooNE \cite{Miniboone} antineutrino  data)
can be perfectly described if we assume that the number of neutrinos with definite masses $\nu_{i}$ is equal to the number of  flavor neutrinos which, as was proved by the LEP experiments, is equal to three.

Neutrino oscillation data are usually analyzed  under the assumption that the flavor neutrino transition probability in vacuum is given by the following standard expression
(see, for example, \cite{BGG})
\begin{equation}\label{oscprobabil}
\mathrm{P}(\nu_{l}\to \nu_{l'})=|\sum_{i}U_{l'i}~e^{-i\frac{\Delta m^{2}_{ki}L}{2E}}~U^{*}_{li}|^{2}=
|\sum_{i\neq k}U_{l'i}~(e^{-i\frac{\Delta m^{2}_{ki}L}{2E}}-1)~U^{*}_{li}+\delta_{l'l}|^{2}.
\end{equation}
Here $L$ is the source-detector distance, $E$ is the neutrino energy, $U$ is the unitary mixing matrix, $\Delta m^2_{ik }=m^2_{k }-m^2_{i}$.

In the case of three-neutrino mixing the unitary $3\times 3$ Pontecorvo-MNS
mixing matrix \cite{BPont,MNS} is usually parameterized by the three Euler angles  $\theta_{12}$,
$\theta_{23}$, $\theta_{13}$ and one $CP$ phase $\delta$. It has the following form
\begin{eqnarray}\label{oscprobabil1}
U=
\left(
\begin{array}{ccc}
  1 & 0& 0\\
  0 & c_{23}  & s_{23} \\
  0 & -s_{23}  &c_{23}  \\
\end{array}%
\right)
\left(%
\begin{array}{ccc}
  c_{13} & 0& s_{13}e^{-i\delta}\\
  0 & 1 & 0\\
  -s_{13}e^{i\delta} & 0 &c_{13}  \\
\end{array}%
\right)
\left(%
\begin{array}{ccc}
  c_{12} & s_{12}& 0\\
  -s_{12} & c_{12} & 0\\
  0 & 0 & 1\\
\end{array}
\right)
\end{eqnarray}
and neutrino transition probabilities in vacuum are characterized by six parameters:
$\theta_{12}$,
$\theta_{23}$, $\theta_{13}$, $\delta$, $\Delta m^2_{12 }$ and $\Delta m^2_{23 }$.

From the analysis of the existing neutrino oscillation data  follows, however,  that two parameters are small:
\begin{equation}\label{oscprobabil2}
\frac{\Delta m^{2}_{12}}{\Delta m^{2}_{23}}\simeq \frac{1}{30},\quad
\sin^{2}\theta_{13}\lesssim 4\cdot 10^{-2}.
\end{equation}
If we neglect the contribution of the small parameters (leading approximation) a rather simple picture of neutrino oscillations has emerged (see, \cite{BGG}). In this approximation neutrino oscillations in the atmospheric (accelerator) region of $\frac{L}{E}$ ($\frac{\Delta m^2_{23 }L }{ 2
E}\gtrsim 1$)  are  two-neutrino $\nu_{\mu}\rightleftarrows \nu_{\tau}$ ($\bar\nu_{\mu}\rightleftarrows \bar\nu_{\tau}$) oscillations. The
$\nu_{\mu}\to \nu_{\mu}$ ($\bar\nu_{\mu}\to \bar\nu_{\mu}$) survival probability is given in this case by the standard two-neutrino expression
\begin{equation}\label{oscprobabil3}
\mathrm{P}(\nu_{\mu} \to\nu_{\mu})=\mathrm{P}(\bar\nu_{\mu} \to\bar\nu_{\mu})=1-\frac{1}{2}~\sin^{2}2\theta_{23}~
(1-\cos\Delta m_{23}^{2}\frac{L}{2E}).
\end{equation}
Neutrino oscillations in the reactor KamLAND region of $\frac{L}{E}$ ($\frac{\Delta m^2_{12 }L }{ 2 E}\gtrsim 1$) are  $\bar\nu_{e}\rightleftarrows  \bar\nu_{\mu,\tau}$ oscillations.
The $\bar\nu_{e}\to \bar\nu_{e}$
survival probability is given in the leading approximation by the expression
\begin{equation}\label{oscprobabil4}
\mathrm{P}(\bar\nu_{e} \to \bar\nu_{e})=1-\frac{1}{2}~\sin^{2}2\theta_{12}~
(1-\cos\Delta m_{12}^{2}\frac{L}{2E}).
\end{equation}
In the leading approximation the probability of the solar neutrinos to survive is given by the two-neutrino $\nu_{e}$ survival probability in  matter which depends on $\Delta m_{12}^{2}$, $\sin^{2}\theta_{12}$ and the electron number density.

The leading approximation gives the dominant contribution to the transition probabilities:
the values of the parameters $\Delta m_{12}^{2}$, $\Delta m_{23}^{2}$, $\sin^{2}\theta_{23}$,
$\sin^{2}\theta_{12}$, which are determined from the two-neutrino and the three-neutrino analysis, are practically the same.

From the three-neutrino analysis of the Super-Kamiokande atmospheric neutrino data
\footnote{In the case of the three-neutrino analysis of the neutrino oscillation data it is important to take into account that neutrino masses are labeled differently
for the normal neutrino mass spectrum (NS) ($m_{1}<m_{2}<m_{3};  \Delta m^2_{12 }\ll \Delta m^2_{23}$) and for the inverted mass spectrum (IS) ($m_{3}<m_{1}<m_{2};  \Delta m^2_{12 }\ll |\Delta m^2_{13}|$).
The smaller and larger neutrino mass squared differences (the same for both neutrino mass spectra) are equal in NS (IS)
$
\Delta m_{12}^{2}~(\Delta m_{12}^{2})$ and
$
\Delta m_{23}^{2}~(|\Delta m_{13}^{2}|),
$
respectively. Thus, we can not use the $\Delta m_{ik}^{2}$ notation in the case of the three-neutrino analysis of the data. One of the possibility is to use for the larger and smaller
neutrino mass-squared differences, independently of the character of the neutrino mass spectrum,  the notations $\Delta m_{A}^{2}$ and $\Delta m_{S}^{2}$.
Notice that for both neutrino mass spectra the elements of the neutrino mixing matrix $U_{li}$ are usually parameterized {\em in the same way} (inspite that they have different meaning).}
\cite{SK}  the following best fit values of the parameters are found
\begin{equation}\label{SK}
\Delta m^{2}_{A}=2.1\cdot 10^{-3}~\rm{eV}^{2},~~\sin^{2} \theta_{23}=0.5,~~\sin^{2} \theta_{13}=0.0.
\end{equation}
In the case of the normal (inverted) neutrino mass spectrum the following 90\% CL limits
were inferred
\begin{equation}\label{SK1}
1.9~ (1.7)\cdot10^{-3}\leq \Delta m^{2}_{A}\leq 2.6~(2.7)\cdot 10^{-3}~\rm{eV}^{2},\quad    0.407\leq \sin^{2} \theta_{23}\leq 0.583.
\end{equation}
For the parameter $\sin^{2} \theta_{13}$ the following bounds were obtained
\begin{equation}\label{SK2}
    \sin^{2} \theta_{13}\leq 4\cdot 10^{-2}~ (9\cdot 10^{-2}).
\end{equation}
The Super-Kamiokande evidence for
neutrino oscillations was confirmed by the accelerator long-baseline
K2K \cite{K2K} and MINOS \cite{Minos} experiments. From the two-neutrino analysis of the MINOS data
 was found
\begin{equation}\label{Minos}
\Delta m_{A}^{2}=(2.43 \pm 0.13)\cdot 10^{-3}~\rm{eV}^{2},\quad
\sin^{2}2 \theta_{23}>0.90.
\end{equation}
From the three-neutrino global analysis of the  KamLAND reactor and solar data
 was obtained \cite{Kamland}
\begin{equation}\label{Kamland1}
\Delta m_{S}^{2}=(7.50^{+0.19}_{-0.20})\cdot 10^{-5}~\rm{eV}^{2},\quad
\tan^{2}\theta_{12}=0.452^{+0.035}_{-0.032}
\end{equation}
For the parameter $\sin^{2}\theta_{13}$  was found
\begin{equation}\label{Kamland2}
\sin^{2}\theta_{13}=0.020^{+0.016}_{-0.018}
\end{equation}
Finally, from the short baseline reactor experiment CHOOZ \cite{Chooz}
the following upper bound was obtained for the parameter $\sin^{2}\theta_{13}$
\begin{equation}\label{Chooz}
\sin^{2}\theta_{13} < 4\cdot 10^{-2}.
\end{equation}
Let us also notice that from the tritium experiments Mainz \cite{Mainz} and Troitsk \cite{Troitsk}
the following upper bounds for the absolute value of neutrino mass  were found
\begin{equation}\label{Mainz}
m_{\beta}\leq 2.3~\mathrm{eV}~(\mathrm{Mainz})\quad
m_{\beta}\leq 2.2~\mathrm{eV}~(\mathrm{Troitsk}) 
\end{equation}

\section{QFT basics of neutrino oscillations}
Our understanding of neutrino oscillations is based on the following assumptions:

I. The Lagrangian of the electroweak interaction is the Standard Model charged current and neutral current Lagrangians. The leptonic part of the CC Lagrangian is given by the following expression
\begin{equation}\label{CC}
\mathcal{L^{CC}_{I}}(x)=-\frac{g}{\sqrt{2}}\sum_{l=e,\mu,\tau}\bar \nu_{l L}(x) \,\gamma_{\alpha}\, l_{L}(x)~ W^{\alpha}(x)+
\mathrm{h.c.}
\end{equation}
Here $g$ is the electroweak constant, $l_{L}(x)=(\frac{1-\gamma_{5}}{2})~l(x)$ is the left-handed component of the leptonic field $l(x)$ and $W^{\alpha}(x)$ is the field of the vector $W^{\pm}$-bosons.

II. The  flavor (active) fields $\nu_{l L}(x)$ in the Lagrangian (\ref{CC}) are mixtures of the fields of neutrinos with definite masses
\begin{equation}\label{mixing}
\nu_{l L}(x)=\sum_{i} U_{l i}\,\nu_{i L}(x).
\end{equation}
Here $\nu_{i }(x)$ is the field of neutrinos  with mass $m_{i}$ and $U$ is a unitary mixing matrix.

The interaction (\ref{CC}) follows from the requirements of the local $SU(2)\times U(1)$
invariance. It was confirmed with high accuracy by numerous experiments on the study of the weak interaction processes.

The existence of the neutrino mixing is confirmed by the neutrino oscillation experiments.
Four neutrino oscillation parameters are known with accuracies in the range (3-10)\%. However, there are many unknowns in the mixing relation (\ref{mixing}).
We do not know
\begin{itemize}
  \item Are neutrinos with definite masses Majorana or Dirac particles?
  \item Is the number of the neutrinos with definite masses equal to the number of flavor neutrinos (three) or larger (in this case sterile neutrinos must exist)?
  \item What is the value of the parameter $\sin^{2}\theta_{13}$?
  \item What is the value of the $CP$ phase $\delta$?
 \item What is the character of the neutrino mass spectrum (normal or inverted)?
\item etc.
\end{itemize}
We believe that the resolution of these problems apparently  will allow to solve the most important problem: {\em What is the origin of small neutrino masses and neutrino mixing?}

Neutrino masses and neutrino mixing are due to {\em the neutrino mass term of the Lagrangian}. According to the Standard Model mass terms of quarks and leptons are generated by the spontaneous violation of
the electroweak symmetry.  The origin of the neutrino mass term at present is unknown. We will consider in this section a general theoretical framework for possible neutrino mass terms. In the next section we will discuss the most popular seesaw mechanism of neutrino mass generation.

Any mass term  is a sum of Lorentz-invariant products of the left-handed and right-handed components of a field. Three left-handed flavor neutrino fields $\nu_{l L}(x)$ must enter into the mass term. Do we need other fields to build the mass term? Generally not, if we assume that the lepton number is not conserved. This was shown for the first time in \cite{GribovP}
.

In fact, it is easy to show that
$(\nu_{l L}(x))^{c}=C \bar \nu_{l L}(x)^{T}$ is the right-handed component ($C\gamma_{\alpha}^{T}C^{-1}=-\gamma_{\alpha}$,~~$C^{T}=-C$).\footnote{The left-handed component satisfies the condition $\gamma_{5}\nu_{L}=-\nu_{L}$. From this relation we have
$\gamma^{T}_{5}\bar \nu^{T}_{L}=\bar\nu^{T}_{L}$. Taking into account that
$C\gamma_{5}^{T}C^{-1}=\gamma_{5}$ we find $\gamma_{5}(\nu_{ L})^{c}=(\nu_{ L})^{c}$.
This relation means that $(\nu_{ L})^{c}$ is the right-handed component.} From
$\nu_{l L}(x)$ and $(\nu_{l L}(x))^{c}$ we can build the following {\bf Majorana mass term}
\begin{equation}\label{Mjmass}
\mathcal{L}^{\mathrm{M}}= -\frac{1}{2}\,
\overline n_{L}\,M^{L} ( n_{L})^{c}
+\mathrm{h.c.}
\end{equation}
Here
\begin{eqnarray}\label{Mjmass1}
n_{L}=\left(
\begin{array}{c}
  \nu_{e L} \\
  \nu_{\mu L}\\
  \nu_{\tau L}
\end{array}
\right)
\end{eqnarray}
and $M^{L}$ is a $3\times3$ complex nondiagonal matrix. Taking into account the Fermi-Dirac statistics of the neutrino field we have
\begin{equation}\label{FDstatistics}
\overline n_{L}M^{L} ( n_{L})^{c}=\overline n_{L}M^{L}C \overline n^{T}_{L}=- \overline n_{L}(M^{L})^{T} C^{T}\overline n^{T} _{L}=\overline n_{L}(M^{L})^{T} ( n_{L})^{c}.
\end{equation}
Thus, {\em $M^{L}$ must be a symmetrical matrix.} A symmetrical, complex matrix can be diagonalized with the help of one unitary matrix:
\begin{equation}\label{Mjmass2}
M^{L}=UmU^{T},
\end{equation}
$U^{\dag}U=1$  and $m$ is a diagonal matrix ($m_{ik}=m_{i}\delta_{ik},~~ m_{i}>0$). From (\ref{Mjmass}) and (\ref{Mjmass2}) we
find
\begin{equation}\label{Mjmass3}
\mathcal{L}^{\mathrm{M}}= -\frac{1}{2}\,\bar \nu m \nu  = -\frac{1}{2}\sum^{3}_{i=1}
m_{i}\bar \nu_{i} \nu_{i}.
\end{equation}
Here
\begin{eqnarray}\label{Mjmass4}
\nu=U^{\dag}n_{L}+ (U^{\dag}n_{L})^{c}=\left(
\begin{array}{c}
  \nu_{1} \\
  \nu_{2}\\
  \nu_{3}
\end{array}
\right).
\end{eqnarray}
From (\ref{Mjmass3}) and (\ref{Mjmass4}) we conclude the following:
\begin{enumerate}
  \item $\nu_{i}(x)$ is the neutrino field with the mass $m_{i}$
  \item The field $\nu_{i}(x)$ satisfies the Majorana condition
\begin{equation}\label{Mjmass5}
\nu^{c}_{i}(x) =\nu_{i}(x).
\end{equation}
From this condition  follows that
\begin{equation}\label{Mjmass6}
\nu_{i}(x)=\int\frac{1}{(2\pi)^{3/2}\sqrt{2p_{0}}}(a_{ir}(p)u^{r}(p)e^{-ipx}+
a^{\dag}_{ir}(p)u^{r}(-p)e^{ipx})d^{3}p.
\end{equation}
Here $a_{ir}(p)$ and $a^{\dag}_{ir}(p) $ are the operators of absorption and creation, respectively, of  a neutrino with mass $m_{i}$, momentum $p$, and helicity $r$.

Thus, if the neutrino field  satisfies the Majorana condition (\ref{Mjmass5})
there is no notion of antineutrino (or, in another words, neutrino and antineutrino are identical). This is connected with the fact that the mass term
 (\ref{Mjmass}) is not invariant under the global gauge transformation
$ \nu_{l L} \to e^{i\Lambda}\nu_{l L}$ i.e.,
 there is no conserved lepton number which would allow to distinguish
 neutrino and antineutrino. Notice that the Majorana mass term (\ref{Mjmass})
 can not be generated in the framework of the SM with a Higgs doublet  (Higgs triplets are necessary).\footnote{
 It is clear from the derivation we presented that the fact that neutrinos with definite masses can be Majorana particles is  based  on the Fermi-Dirac property of neutrino fields. If we assume that neutrino fields are Bose-Einstein fields (this possibility was discussed  in \cite{Dolgov}) then  neutrinos with definite masses can not be Majorana particles. We can see this considering the mass term for a Majorana particle with a mass $m$. We have $\mathcal{L}^{\mathrm{M}}= -\frac{1}{2}\,m
\bar \nu \nu$, where $ \nu=\nu^{c}=C\bar\nu^{T}$. From this last relation we obtain $\overline\nu^{c}=-\nu^{T}C^{-1}$. Now we have $\bar \nu \nu=  \overline\nu^{c}\nu^{c}=-\nu^{T}C^{-1}C\bar\nu^{T}=-(\nu^{T}\bar\nu^{T})^{T}=
+\bar \nu \nu (\mathrm{Fermi})=-\bar \nu \nu (\mathrm{Bose})$. Thus, for a  "bosonic neutrino" $\bar \nu \nu \equiv 0$.}

\item From (\ref{Mjmass4}) follows that the flavor field $\nu_{lL}(x)$ is a  mixture of three Majorana fields $\nu_{iL}(x)$:
\begin{equation}\label{Mjmass7}
\nu_{lL}(x)=\sum_{i}U_{li}\nu_{iL}(x).
\end{equation}
\end{enumerate}
We will assume now that not only flavor fields $\nu_{lL}(x)$, components of the lepton doublets, but also singlet (sterile) fields $\nu_{lR}(x)$ enter into the neutrino mass term. There can be two different  mass terms in this case. We will consider first {\bf the Dirac mass term}
\begin{equation}\label{Dmassterm}
\mathcal{L}^{\mathrm{D}}(x)=- \sum_{l'l}\bar\nu_{ l'L}(x)\,
M_{l'l}^{\mathrm{D}}\,\nu_{ lR}(x)
+\rm{h.c.},
\end{equation}
where $M^{\mathrm{D}}$ is a $3\times3$ complex matrix.

The matrix $M^{\mathrm{D}}$ can be diagonalized by a biunitary transformation. We have
\begin{equation}\label{Dmassterm1}
M^{\mathrm{D}}=U^{\dag}m V,
\end{equation}
where $U$ and $V$ are unitary matrices and
$m_{ik}=m_{i}\delta_{ik}$. From (\ref{Dmassterm}) and (\ref{Dmassterm1}) we find
\begin{equation}\label{Dmassterm2}
 \mathcal{L}^{\mathrm{D}}(x)=
   \sum_{i=1}^{3}  m_{i}\,\bar\nu_{i}(x)\,\nu_{i}(x),
\end{equation}
where
\begin{equation}\label{Dmassterm3}
\nu_{i}(x)=\sum_{l}U^{\dag}_{il}\nu_{lL }(x)+\sum_{l}V_{il}\nu_{lR }(x).
\end{equation}
From  (\ref{Dmassterm2}) and (\ref{Dmassterm3}) we can make the following conclusions:
\begin{enumerate}
  \item The field $\nu_{i}(x)$ is the field of neutrinos with the mass $m_{i}$.
 \item The flavor fields $\nu_{lL}(x)$ are connected with the left-handed components of the fields of neutrinos with definite masses by the mixing relation
 \begin{equation}\label{Dmassterm4}
 \nu_{lL }(x)=\sum^{3}_{i=1}U_{li}~\nu_{i L}(x).
 \end{equation}
\end{enumerate}
The Lagrangian with the neutrino mass term  (\ref{Dmassterm}) is invariant under the global phase transformations
\begin{equation}\label{Dmassterm5}
\nu_{i}(x)\to e^{i\,\Lambda}\nu_{i}(x),~ l(x) \to e^{i\,\Lambda}\,l(x),~ q(x) \to
q(x),
\end{equation}
where $\Lambda$ is an arbitrary constant.
From the invariance under the transformations (\ref{Dmassterm2}) follows that
the total lepton number $L$, the same for $e$, $\mu$  and $\tau$, is conserved. The field $\nu_{i}(x)$ is the four-component Dirac field of neutrinos and antineutrinos with the same mass $m_{i}$ and different lepton numbers ($L(\nu_{i})=1, L(\bar\nu_{i})=-1$).\footnote{From (\ref{Dmassterm4}) and (\ref{Dmassterm5})
we find $\nu_{lL}(x)\to e^{i\,\Lambda}\nu_{lL}(x)$. Thus, for the flavor neutrinos we have: $L(\nu_{l})=1, L(\bar\nu_{l})=-1$.}

The Dirac neutrino mass term can be generated by the standard Higgs mechanism, which is responsible for the generation of the masses of quarks and leptons. However, this mechanism can not explain the smallness of the neutrino masses with respect to the masses of quarks and leptons.

There is no fundamental principle which requires the conservation of the lepton number $L$. The barion asymmetry of the Universe signifies that the barion number is violated. It is natural to assume that in some interaction the lepton number is also violated. If this interaction is relevant for the generation of the neutrino masses, the neutrino mass term will violate the lepton number. The most general neutrino mass term which violates the lepton number is {\bf the Dirac and Majorana mass term}
\begin{equation}\label{DMmass}
\mathcal{L}^{\mathrm{D+M}}= -\frac{1}{2}\,
   \sum_{l'l}\overline \nu_{l'L}\,M^{L}_{l'l} ( \nu_{lL})^{c}-
 \sum_{l'l}\overline\nu_{l'L}\,M^{\mathrm{D}}_{l'l} \, \nu_{lR}
-\frac{1}{2}\, \sum_{l'l}\overline{(\nu_{l'R})^{c}}\,M^{R}_{l'l}  \nu_{lR}
+\mathrm{h.c.}
\end{equation}
Here $M^{L}$ and $M^{R}$ are complex, symmetrical $3\times3$ matrices and  $M^{\mathrm{D}}$ is a complex  $3\times3$ matrix.
After the diagonalization of this mass term we find
\begin{equation}\label{DMmass1}
\nu_{l L}=\,\sum^{6}_{i=1}U_{l i}\,\nu_{i L},\qquad(\nu_{l R})^{c}=
\,\sum^{6}_{i=1}U_{\bar l i}\,\nu_{i L}~~l=e,\mu,\tau
\end{equation}
and
\begin{equation}\label{DMmass2}
\mathcal{L}^{\mathrm{D+M}}(x)=-\frac{1}{2}\sum_{i=1}^{6}  m_{i}\,\bar\nu_{i}(x)\,\nu_{i}(x).
\end{equation}
Here the field $\nu_{i}(x)$
satisfies the  condition
\begin{equation}\label{DMmass3}
\nu_{i}(x)=\nu^{c}_{i}(x)=C\bar\nu_{i}^{T}(x)
\end{equation}
and $U$ in (\ref{DMmass1}) is a $6\times 6$ unitary mixing matrix. From (\ref{DMmass2}) and (\ref{DMmass3}) follows that the field $\nu_{i}(x)$ is a field of Majorana particles with mass $m_{i}$.

From the consideration of the Dirac and Majorana mass term we can conclude that
the number of the massive neutrinos can be larger than the number of the flavor neutrinos (three).
 Let us write  in general
 \begin{equation}\label{DMmass4}
\nu_{l L}=\sum^{3+n_{s}}_{i=1}U_{l i}\,\nu_{i L}\quad l=e,\mu,\tau
 \end{equation}
and
\begin{equation}\label{DMmass5}
\nu_{sL}=
\sum^{3+n_{s}}_{i=1}U_{s i}\,\nu_{i L}\quad s=s_{1},...s_{n_{s}}
\end{equation}
Thus, we assumed that the three flavor neutrino fields $\nu_{lL}$ are mixtures of the left-handed components of $3+n_{s}$ massive fields. This means that other
 $n_{s}$ mixtures of  left-handed components of the same $3+n_{s}$ massive fields must exist. We denoted them $ \nu_{sL}$. The fields $ \nu_{sL}$ do not enter into the standard weak interaction Lagrangian and are called {\ sterile fields}.

All flavor neutrinos ($\nu_{e}, \nu_{\mu},\nu_{\tau}$) were observed in experiments. Sterile neutrinos $\nu_{s}$ can not be produced in weak processes.
There are two ways to reveal the existence of the sterile neutrinos.

I. If neutrinos are detected via the observation of NC processes
the sum of the probabilities of the transitions into all flavor neutrinos
 $\sum_{l'=e,\mu.\tau} \mathrm{P}(\nu_{l} \to \nu_{l'})$
will be measured. If there are no transitions into sterile neutrinos $\sum_{l'=e,\mu.\tau} \mathrm{P}(\nu_{l} \to \nu_{l'})=1$ and
no oscillations will be observed. If there are transitions into sterile neutrinos $\sum_{l'=e,\mu.\tau} \mathrm{P}(\nu_{l} \to \nu_{l'})=1-\sum_{s} \mathrm{P}(\nu_{l} \to \nu_{s})$ and neutrino oscillations can be observed.

II. Neutrino oscillations with two neutrino mass-squared differences $\Delta m^{2}_{A}$ and $\Delta m^{2}_{S}$ were observed in different experiments. If oscillations with additional mass-squared difference(s) will be measured this will be the proof of the existence of sterile neutrino(s).

During many years the LSND indication \cite{LSND} in favor of the  $\bar \nu_{\mu}\to \bar \nu_{e}$ transition with $\Delta m^{2}\simeq 1 ~\mathrm{eV}^{2}$)($\Delta m^{2}\gg\Delta m^{2}_{A,S}$). In the MiniBooNE experiment this indication was checked.
In the channel $\nu_{\mu}\to \nu_{e}$ the LSND result was not confirmed \cite{Miniboone1}.
In the channel $\bar \nu_{\mu}\to \bar \nu_{e}$ some indication in favor of neutrino oscillations, compatible with the LSND result, was obtained \cite{Miniboone}. Further experiments are necessary in order to test the idea of a possible existence of the sterile neutrinos.

\section{On the seesaw mechanism of neutrino mass generation}

Neutrino masses are many orders of magnitude smaller than masses of quarks and leptons. Let us consider, for example, the masses of the particles of the third family. We have
\begin{equation}\label{seesaw}
m_{t}\simeq 1.7 \cdot 10^{2}~\mathrm{GeV},~~m_{b}\simeq 4.7 ~\mathrm{GeV},~~
{\bf m_{3}\leq 2.3~10^{-9}~\mathrm{GeV}},~~m_{\tau}\simeq 1.8~\mathrm{GeV}.
\end{equation}
From these values we can conclude that it is very unlikely that the masses of quarks, leptons and neutrinos are of the same origin. We believe that the masses of the quarks and leptons are due to the standard Higgs mechanism. {\em For neutrino masses a new (or additional) mechanism is needed.}
We will discuss here the most popular  seesaw mechanism of the  generation of small neutrino masses \cite{seesaw}. There are different versions of this mechanism. We will discuss first the mechanism which is based on the Dirac and Majorana mass term.

For illustration let us consider  the Dirac and Majorana mass term in the simplest case of one generation. We have
\begin{equation}\label{seesaw1}
 \mathcal{L}^{\mathrm{D+M}}=-\frac{1}{2}~m_{L} \overline \nu_{L}
(\nu_{L})^{c}-m_{D} \overline \nu_{L} \nu_{R}-\frac{1}{2}~m_{R}
\overline{(\nu_{R})^{c}} \nu_{R} +\mathrm{h.c.},
\end{equation}
where   $m_{L,R}$ and $m_{D}$ are real parameters. The mass term (\ref{seesaw1}) can be easily diagonalized. We have
\begin{equation}\label{seesaw2}
\mathcal{L}^{\mathrm{D+M}}=-\frac{1}{2}\, \sum_{i=1,2} m_{i}\,\overline \nu_{i}\,\nu_{i},    \end{equation}
where $\nu_{1,2}$ are Majorana fields with masses $m_{1,2}$ and
\begin{equation}\label{seesaw3}
\nu_{L}= \cos\theta~\nu_{1 L} + \sin\theta~\nu_{2 L},\quad
(\nu_{R})^{c}=- \sin\theta ~\nu_{1 L} + \cos\theta~
\nu_{2 L}.
\end{equation}
The neutrino masses $m_{1,2}$ and the mixing angle $\theta$ are connected with the parameters
$m_{L,R}$ and $m_{R}$ by the following relations
\begin{equation}\label{seesaw4}
m_{1,2}= \frac{1}{2}~\left|(m_{R}+m_{L})  \mp
\sqrt{(m_{R}-m_{L})^{2} + 4\,m_{D}^{2}}\right|
\end{equation}
and
\begin{equation}\label{seesaw5}
\tan 2\,\theta=\frac{2m_{D}}{m_{R}-m_{L}}.
\end{equation}
We will assume now that
\begin{enumerate}
  \item There is no left-handed Majorana mass term in the Lagrangian, i.e.
  $m_{L}=0$
\item The Dirac mass term is generated by the Standard Higgs mechanism, i.e. $m_{D}$ is of the order of a mass of a quark or a lepton.
\item A new mechanism generates a right-handed Majorana mass term. This term does not conserve the lepton number. We assume that the lepton number is violated at a scale which is much larger than the electroweak scale, i.e. that
$m_{R}\equiv M_{R}\gg m_{D}$.
\end{enumerate}
From (\ref{seesaw4}) and (\ref{seesaw5}) we obtain \footnote{If $m_{D}\simeq m_{t}\simeq 170 $ GeV and $m_{1}\simeq 5\cdot 10^{-2}$, we find   $M_{R}\simeq 10^{15}$ GeV.}
\begin{equation}\label{seesaw6}
m_{1}\simeq \frac{m^{2}_{D}}{M_{R}}\ll m_{D},\quad m_{2} \simeq M_{R}\gg m_{D},\quad
\theta \simeq \frac{m_{D}}{M_{R}}\ll 1.
\end{equation}
Thus, in the example we have considered there are two masses in the Majorana mass spectrum:  very light (neutrino mass) and very heavy (mass of a new particle). The mixing angle is tiny.

In the case of three families the seesaw matrix has the form
\begin{eqnarray}\label{seesaw6}
 M=\left(\begin{array}{cc}
0&m_{D}\\
m^{T}_{D}&M_{R}\end{array}\right)
\end{eqnarray}
Here $m_{D}$ and $M_{R}=M^{T}_{R}$ are 3$\times$3 matrices and  $M_{R}\gg m_{D}$.
The matrix $M$ can be presented in block-diagonal form by the unitary transformation
\begin{eqnarray}\label{seesaw7}
 U^{T}\,M\,U =\left(\begin{array}{cc}
-m_{D}M_{R}^{-1}m^{T}_{D}&0\\
0&M_{R}\end{array}\right).
\end{eqnarray}
The $3\times3$ Majorana  mass matrix is given by
\begin{equation}\label{seesaw8}
m_{\nu}=-m_{D}~M_{R}^{-1}~m^{T}_{D}.
\end{equation}
There are many parameters in the matrix $m_{\nu}$. The large  denominator
$M_{R}^{-1}$ ensures, however, the smallness of the neutrino masses with respect to the masses of leptons and quarks.
We can make the following conclusions:
\begin{enumerate}
 \item In the seesaw approach neutrinos with definite masses are Majorana particles.
  \item The smallness of neutrino masses is due to a right-handed Majorana mass term which violates the lepton number at a large scale. The suppression factors which provide  the smallness of neutrino masses is characterized
 by the ratio of the electroweak scale  and the scale of the violation of the lepton number.
 \item Heavy Majorana particles, partners of light Majorana neutrinos, must exist.
\end{enumerate}
We have discussed the seesaw idea in terms of the Dirac and Majorana mass term.
 The same idea can be realized in another way. Let us assume that there exist heavy Majorana fermions $N_{i}$, singlets of the $SU(2)\times U(1)$ group, which have the following $SU(2)\times U(1)$ invariant Yukawa interaction with leptons and standard Higgs bosons
\begin{equation}\label{heavyH}
\mathcal{L}=\sqrt{2}\sum_{i,l}Y_{il}\overline  L_{lL} N_{iR} \tilde{\phi }+
\mathrm{h.c.}.
\end{equation}

Here $Y_{il}$ are dimensionless constants and
\begin{eqnarray}\label{heavyH1}
L_{lL}
=
\left(
\begin{array}{c}
\nu_{lL}
\\
l_L
\end{array}
\right)
\qquad
\phi
=
\left(
\begin{array}{c}
\phi^{(+)}
\\
\phi^{(0)}
\end{array}
\right)
\end{eqnarray}
are lepton and Higgs doublets and $\tilde{\phi}=i\tau_{2}\phi^{*}$ is the conjugated Higgs doublet. We assume that $M_{i}\gg v$ where $M_{i}$ is the mass of the Majorana fermion $N_{i}$ and  $v\simeq 246$ GeV is the Higgs vacuum expectation value.
It is obvious that the Lagrangian (\ref{heavyH}) does not conserve $L$.  For the processes with virtual $N_{i}$ at $Q^{2}\ll M^{2}_{i}$ the interaction (\ref{heavyH})
generates  a non-renormalizable effective Lagrangian \cite{Weinberg}
\begin{equation}\label{heavyH2}
\mathcal{L}_{\rm{eff}}=-
\sum_{l',l,i}\,\overline L_{l'L}\tilde{\phi }Y_{il'}\frac{1}{M_{i}}Y_{il}C
\tilde{\phi }^{T}(\overline L_{lL})^{T}
+
\mathrm{h.c.}
\end{equation}
If we put
\begin{eqnarray}\label{heavyH3}
\tilde{\phi} =\left(
  \begin{array}{c}
 \frac{ v +H}{\sqrt{2}}   \\
   0\\
  \end{array}
\right)
\end{eqnarray}
($H$ is the Higgs field) the electroweak symmetry will be spontaneously broken and
from (\ref{heavyH2}) we obtain the left-handed Majorana mass term
\begin{equation}\label{heavyH4}
\mathcal{L}^{\mathrm{M}}=-
\frac{1}{2}
\,\sum_{l'l}
\overline \nu_{l'L}~
M^{L}_{l'l}~
(\nu_{lL})^{c}
+
\mathrm{h.c.},
\end{equation}
where
\begin{equation}\label{heavyH5}
M^{L}=Y^{T}\frac{v^{2}}{M}Y
\end{equation}
is the seesaw mass matrix.\footnote{The model we have discussed is usually called the seesaw type I model. The model based on the interaction of lepton pairs and the Higgs pair with heavy scalar  triplet bosons is called the seesaw type II model and the model based on the interaction of lepton-Higgs  pairs with heavy Majorana triplet fermions is called the seesaw type III model.} The $CP$ violating decays of heavy Majorana fermions $N_{i}$ in the early Universe are considered as a possible source of the barion asymmetry of the Universe (see \cite{Nir}).

\section{On the nature of neutrino oscillations}

\subsection{Introduction}

A lot of debates on the nature of neutrino oscillations can be found in the literature (see recent papers \cite{Oscillations}). We will discuss here this problem.

From our point of view {\em the Heisenberg uncertainty  relation and the time-energy uncertainty relation  are crucial for the phenomenon of  neutrino oscillations}. Uncertainty relations in Quantum Theory are based on the inequality
 \begin{equation}\label{uncertain}
\Delta A~\Delta B
\geq \frac{1}{2}|\langle a|[A,B]|a\rangle|,
 \end{equation}
which can be easily derived from the Cauchy inequality.
Here $A$ and $B$ are hermitian operators, $|a\rangle$ is any state and
$\Delta A=\sqrt{\overline {A^{2}}-\overline {A }^{2}}$ is the  standard deviation.
For example, for operators $p$ and $q$ which satisfy the relation $[p,q]=\frac{1}{i}$  we obtain from (\ref{uncertain}) the Heisenberg uncertainly relation $\Delta p~\Delta q \geq \frac{1}{2}$.

There exist different derivations of the time-energy uncertainty relation
\begin{equation}\label{uncertain3}
\Delta E ~\Delta t \geq 1 .
\end{equation}
and different interpretations of the quantities which enter into this relation (see, for example,  \cite{Busch}). Mandelstam and Tamm \cite{MandTamm} derived the relation (\ref{uncertain3}) from inequality (\ref{uncertain}) and the evolution equation
\begin{equation}\label{uncertain1}
i\frac{d O(t)}{d t}=[O(t),H]
\end{equation}
for an operator $O(t)$ in the Heisenberg representation ($H$ is the total Hamiltonian).
From (\ref{uncertain}) and (\ref{uncertain1})
we have
\begin{equation}\label{uncertain2}
\Delta E~\Delta O(t)\geq\frac{1}{2}|\frac{d }{d t}\overline{ O}(t)|.
\end{equation}
For stationary states  (\ref{uncertain2}) is identically satisfied. Nontrivial constraints can be obtained only in the
case of nonstationary states. In \cite{MandTamm}
the  time-energy  uncertainty relation (\ref{uncertain3}) was derived in which $\Delta E$ is the uncertainty of the energy of the system and
{\em $\Delta t$ is the time interval during which the state of the system is significantly changed.}

\subsection{Flavor neutrino states}

We will consider the neutrino production. Neutrinos are produced in weak decays and reactions. Let us consider (in the lab. system) the decay \cite{BilGiun}
\begin{equation}\label{decay}
 a\to b +l^{+}+\mathrm{neutrino}
\end{equation}
where $a$ and $b$ are some hadrons.  The sum of the states of the final particles is given by
\begin{equation}\label{decay1}
|f\rangle=\sum_{i}|b~\rangle|l^{+}\rangle|\nu_{i}\rangle
\langle b ~l^{+}\nu_{i}|S|a\rangle,
\end{equation}
where $\langle b ~l^{+}\nu_{i}|S|a\rangle$ is the matrix element of the transition
$a\to b +l^{+}+\nu_{i}$ where $\nu_{i}$ is the neutrino with mass $m_{i}$. We assume, as usual, that initial and final particles have definite momenta. Momenta of neutrinos with mass $m_{i}$ will be denoted by $p_{i}$.

Neutrinos $\nu_{i}$ differ only by their masses. If  masses of neutrinos are the same, their momenta will be equal. Taking into account that neutrino masses are much smaller than neutrino momenta we have
\begin{equation}\label{decay2}
p_{i}\simeq p+a\frac{\Delta m_{1i}^{2}}{2E},
\end{equation}
where  $p$ is the momentum of the lightest neutrino,
$E\simeq p$ is the neutrino energy and $|a|\lesssim 1$ is a constant.
For the difference of the neutrino momenta we have
\begin{equation}\label{decay3}
|p_{i}-p_{k}|\lesssim \frac{|\Delta m^{2}_{ik}|}{2E}=\frac{1}{l_{ik}}
\end{equation}
For reactor and atmospheric (accelerator) neutrinos we find, respectively,
\begin{equation}\label{decay5}
l_{12}\simeq 15~\mathrm{km}, \quad l_{23}\simeq 200~\mathrm{km}.
\end{equation}
For the uncertainty of the neutrino momentum we have
\begin{equation}\label{decay6}
 (\Delta p )_{QM}\simeq \frac{1}{d} ~,
\end{equation}
where $d$ characterizes  the quantum-mechanical size of the source.

Because the macroscopic length $l_{ik}$ is much larger than the microscopic
quantum-mechanical size of the source we have
\begin{equation}\label{decay7}
|p_{i}-p_{k}|\ll (\Delta p )_{QM}.
\end{equation}
Thus, due to the uncertainty relation it is impossible to resolve the momenta of neutrinos with different masses. Because $E_{i}\simeq p_{i}(1 + \frac{m^{2}_{i}}{2E^{2}})$ and $\frac{m^{2}_{i}}{2E^{2}}\leq 10^{-13}$, energies of neutrinos with different masses also can not be resolved.

Let us consider the lepton part of the matrix element
$\langle b ~l^{+}\nu_{i}|S|a\rangle$. Taking into account  inequality (\ref{decay3}), we have
\begin{equation}\label{decay8}
U^{*}_{li}~\bar u_{L}(p_{i})\gamma_{\alpha}u(-p_{l})\simeq
U^{*}_{li}\bar u_{L}(p)\gamma_{\alpha}u(-p_{l}),
\end{equation}
where $p_{l}$ is the momentum of $l^{+}$. For the total matrix element we have
\begin{equation}\label{decay9}
\langle b ~l^{+}\nu_{i}|S|a\rangle\simeq U^{*}_{li}~\langle b ~l^{+}\nu_{l}|S|a\rangle_{SM},
\end{equation}
where $\langle b ~l^{+}\nu_{l}|S|a\rangle_{SM}$ is the Standard Model matrix element of the emission of the flavor neutrino $\nu_{l}$\footnote{By definition the flavor neutrino $\nu_{l}$ is a particle which is emitted in a weak process together with $l^{+}$ and the flavor antineutrino $\bar \nu_{l}$ is a particle which is emitted together with $l^{-}$.}
with momentum $p$ and $l^{+}$ in the process
\begin{equation}\label{decay10}
 a\to b +l^{+}+\nu_{l}.
\end{equation}
From (\ref{decay1}) and (\ref{decay9}) we find
\begin{equation}\label{decay11}
|f\rangle=|b~\rangle|l^{+}\rangle|\nu_{l}\rangle
\langle b ~l^{+}\nu_{l}|S|a\rangle_{SM}.
\end{equation}
Here
\begin{equation}\label{decay12}
|\nu_{l}\rangle=\sum^{3}_{i=1}U^{*}_{li}~|\nu_{i}\rangle \quad (l=e,\mu,\tau)
\end{equation}
is {\em the state of the flavor neutrino $\nu_{l}$}.
Thus, due to the smallness of the neutrino mass-squared differences  and the uncertainty relation it is impossible to say which massive neutrino is emitted in a weak process.  This is the reason why {\em a coherent superposition of states of neutrinos with different masses is produced.} Let us stress that
\begin{itemize}
  \item Flavor neutrino states do not depend on the production process (for example, $\nu_{e}$'s produced in $\mu$-decay and in $\beta$-decay are the same particles).
  \item It is natural to assume that flavor states are characterized by the momentum (if there are no special conditions of neutrino production).
\item Flavor states are orthogonal and normalized
\begin{equation}\label{decay13}
\langle \nu_{l'}|\nu_{l}\rangle=\delta_{l'l}.
\end{equation}
\end{itemize}

\subsection{Schr\"odinger evolution of flavor neutrino states}

The evolution equation for states in QFT is the Schr\"odinger equation
\begin{equation}\label{Schrod}
i\frac{\partial~ |\Psi(t)\rangle}{\partial t}=H~|\Psi(t)\rangle,
\end{equation}
where $H$ is the total Hamiltonian. The general solution of this equation is given by
\begin{equation}\label{Schrod1}
 |\Psi(t)\rangle=e^{-iHt}|\Psi(0)\rangle.
\end{equation}
If at $t=0$ in a CC weak process $\nu_{l}$ is  produced,
 for the neutrino state we have  at the time $t$
\begin{equation}\label{Schrod2}
|\nu_{l}\rangle_{t}=
\sum_{i}|\nu_{i}\rangle e^{-iE_{i}t}~U^{*}_{li},
\end{equation}
where $E_{i}=\sqrt{p^{2}+m^{2}_{i}}$.

Neutrinos are detected via observation of weak processes. Let us consider the transition
\begin{equation}\label{detect}
\nu_{i}+N\to l' +X.
\end{equation}
For the matrix element we have
\begin{equation}\label{detect1}
\langle l' ~X|S|\nu_{i}~N \rangle\simeq
\langle l' ~X|S|\nu_{l'}~N\rangle_{SM}~U_{l'i},
\end{equation}
where $\langle l' ~X|S|\nu_{l'}~N\rangle_{SM}$ is the SM matrix element of the process
\begin{equation}\label{detect2}
\nu_{l'}+N\to l' +X.
\end{equation}
From (\ref{decay11}), (\ref{Schrod2}) and (\ref{detect1}) follows that to the chain  of processes $a\to b +l^{+}+\nu_{l}$,~~ $\nu_{l}\to \nu_{l'}$,
~~$\nu_{l'}+N\to l'+X$ corresponds the {\em factorized product of amplitudes}
\begin{equation}\label{detect3}
\langle l' ~X|S|\nu_{l'}~N\rangle_{SM}~
\left(\sum_{i}U_{l'i}~e^{-iE_{i}t}~U^{*}_{li}\right)~\langle b ~l^{+}\nu_{l}|S|a\rangle_{SM}.
\end{equation}
Only the amplitude of the transition $\nu_{l}\to \nu_{l'}$
\begin{equation}\label{detect4}
\mathcal{A}(\nu_{l}\to \nu_{l'})=\sum_{i}U_{l'i}~e^{-iE_{i}t}~U^{*}_{li}
\end{equation}
depends on the properties of massive neutrinos (mass-squared differences and mixing angles). The matrix elements of the neutrino production and detection do not depend on any characteristics of individual massive neutrinos. They are given by the Standard Model. Let us stress that the important property of the factorization (\ref{detect3}) is based on the Heisenberg uncertainty relation.

For the probability of the transition $\nu_{l}\to \nu_{l'}$ we have
\begin{equation}\label{probabil}
P(\nu_{l}\to \nu_{l'})=|\sum_{i}U_{l'i}~e^{-i(E_{i}-E_{k})t}~U^{*}_{li}|^{2}=
|\sum_{i\neq k}U_{l'i}~(e^{-i(E_{i}-E_{k})t}-1)~U^{*}_{li}+\delta_{l'l}|^{2}.
\end{equation}
From this expression it is obvious that neutrino oscillations can be observed if the condition\footnote{This is a necessary condition for the observation of the oscillations. Another condition: relatively large mixing angles.}
\begin{equation}\label{probabil1}
|E_{i}-E_{k}|~t \gtrsim 1,\quad (i\neq k)
\end{equation}
is satisfied. This inequality is the Mandelstam-Tamm time-energy uncertainty relation. According to this relation a change of the flavor neutrino state in time requires energy uncertainty (nonstationary state). The time interval required for a significant change of the flavor neutrino state (oscillations) is given by $t \simeq \frac{1}{|E_{i}-E_{k}|}$.

The inequality (\ref{probabil1}) can be interpreted in another way:
in order to resolve a small energy difference  $|E_{i}-E_{k}|\simeq \frac{|\Delta m_{ik}^{2}|}{2E}$ we need a macroscopically large time interval $t \gtrsim \frac{1}{|E_{i}-E_{k}|}$. This corresponds to another interpretation of the time-energy uncertainty relation (see \cite{Fock}).

The time $t$ in the equation (\ref{Schrod}) is a parameter which in our case describes the propagation of the neutrino signal. For the ultrarelativistic neutrino we have $t\simeq L$, where $L$ is the distance between the neutrino source and the detector. Taking into account this relation and the relation $E_{i}-E_{k}=\frac{\Delta m^{2}_{ki}}{2E}$ from
(\ref{probabil}) we obtain the standard expression (\ref{oscprobabil}) for the neutrino transition probability.

\subsection{On other approaches to neutrino oscillations}

We will now briefly describe other approaches to neutrino oscillations which were considered in the literature. We will start with the following remark.  In many papers
(see, for example, \cite{Levy}) the covariant operator $e^{-iPx}$ ($P^{{\alpha}}$ is the operator of the total momentum and $x^{{\alpha}}=(t,\vec{x})$ is the space-time point) is applied to the mixed
flavor neutrino states (\ref{decay12}). If we assume that
at point $x=0$ the flavor neutrino $\nu_{l}$ is produced, we have for the neutrino state at the point $x$  in this case
\begin{equation}\label{xevolution}
|\nu_{l}\rangle_{x}=e^{-iPx}|\nu_{l}\rangle=
\sum_{i}e^{-ip_{i}x}U^{*}_{li}|\nu_{i}\rangle=\sum_{l'}|\nu_{l'}\rangle
\left(\sum_{i}U_{l'i}e^{-ip_{i}x}U^{*}_{li}\right)
\end{equation}
For the probability of the transition $\nu_{l}\to \nu_{l'}$ we find the following expression
\begin{equation}\label{xevolution1}
P(\nu_{l}\to \nu_{l'})=
|\sum_{i}U_{l'i}e^{-ip_{i}x}U^{*}_{li}|^{2}=\sum_{i\neq k}U_{l'i}(e^{-i(p_{i}-p_{k})x}-1)U^{*}_{li}+\delta_{l'l}|^{2}
\end{equation}
Let us assume that
$\vec{p}_{i}=p_{i}\vec{k}$, where $\vec{k}$ is the unit vector. For the phase difference we have
\begin{equation}\label{xevolution2}
(p_{i}-p_{k})x = (E_{i}-E_{k})t-(p_{i}-p_{k})L\simeq \frac{\Delta m^{2}_{ki}L}{2E}+(E_{i}-E_{k})(t-L).
\end{equation}
Taking into account that for the ultrarelativistic neutrinos $t=L$ we obtain from (\ref{xevolution1}) and (\ref{xevolution2})  the standard expression
(\ref{oscprobabil}) for the neutrino transition probability.\footnote{
In the approach based on the Schr\"odinger equation, the
phase difference is equal to $\frac{\Delta m^{2}_{ki}}{2E}L$ if the flavor state possesses  one momentum. We came here to the same result for the phase difference because the neutrino energies in space and time terms are canceled due to  the relation $t\simeq L$ .}
Nevertheless
the presented "derivation" of the transition probability is wrong.
There are two reasons for that:
\begin{itemize}
  \item The operator $e^{-iPx}$ is the operator of the evolution of {\em fields, but not states.} In fact, from the translational invariance for a field operator $\psi(x)$ we have
\begin{equation}\label{translation}
i~\partial_{\alpha}\psi(x)=[\psi(x), P_{\alpha}].
\end{equation}
The general solution of this equation has the form
\begin{equation}\label{translation1}
\psi(x)=e^{iPx}~\psi(0)~e^{-iPx}.
\end{equation}
This equation means that $e^{-iPx}$
is the operator of evolution of fields.

\item The flavor state $|\nu_{l}\rangle$, given by equation (\ref{decay12}), which
describes the mixture of states with definite momenta, can not depend on $x$. In fact, we have
\begin{equation}\label{translation2}
|\nu_{i}\rangle=c_{-1}^{\dag}(p_{i})|0\rangle,
\end{equation}
where $|0\rangle$ is the vacuum state and $c_{-1}^{\dag}(p_{i})$ is the creation operator of a neutrino with momentum $p_{i}$,
mass $m_{i}$ and helicity equal to -1. This operator can not depend on $x$.
\end{itemize}
The expression (\ref{xevolution1}) for the transition probability, in which neutrino mass states evolve in space and time, can be treated only in the framework of {\bf relativistic quantum mechanics.} In this case the wave function of a flavor neutrino $\nu_{l}$, produced in a CC process, is the superposition
\begin{equation}\label{QM}
\psi_{\nu_{l}}(\vec{x},t) =\sum_{i}U^{*}_{li}~\psi_{i}(\vec{x},t),
\end{equation}
where
\begin{equation}\label{QM1}
\psi_{i}(\vec{x},t)=e^{i(\vec{p}_{i}\vec{x}-E_{i}t)}u^{(-1)}(p_{i})
\end{equation}
is the solution of the Dirac equation
\begin{equation}\label{QM2}
i\gamma^{\alpha}\partial_{\alpha}\psi_{i}(\vec{x},t)=m_{i}\psi_{i}(\vec{x},t).
\end{equation}
From (\ref{QM}) and (\ref{QM1}) we find that the normalized probability of the transition $\nu_{l}\to \nu_{l'}$ is given
by the expression (\ref{xevolution1}) in which $p_{i}x = E_{i}t-p_{i} L$ is the change of the phase of the plane wave at the distance $L$ after the time $t$. From (\ref{xevolution1}), as we have shown before,  the standard expression for the transition probability follows.

Let us stress that
\begin{itemize}
  \item in the approach  based on the relativistic
quantum mechanics the notion of flavor neutrino states does not appear.
  \item the "mixed" wave function $\psi_{\nu_{l}}(\vec{x},t)$ does not satisfy the Dirac equation:\footnote{If any wave function of a particle with spin 1/2  must satisfy the Dirac equation, QM is not the appropriate framework for neutrino oscillations.}
\begin{equation}\label{QM3}
i\gamma^{\alpha}\partial_{\alpha}\psi_{\nu_{l}}(\vec{x},t)=
\sum_{i}U^{*}_{li}m_{i}\psi_{i}(\vec{x},t)\neq m~ \psi_{\nu_{l}}(\vec{x},t)
\end{equation}
\item in order to obtain  from the probability
(\ref{xevolution1}), which depends on $x$ and $t$, the standard transition probability we need to assume that
\begin{equation}\label{QM4}
L\simeq t.
\end{equation}
\end{itemize}
We will now briefly discuss   the {\bf wave packet approach to the neutrino oscillations}
(see \cite{Giunti} and references therein). We will see that this approach provides the equality (\ref{QM4}).

 Let us take into account the distribution of momenta of the initial neutrinos determined by the  uncertainty relation.  For the $\nu_{l}\to \nu_{l'}$
transition amplitude we have in this case
\begin{equation}\label{wavepack}
\mathcal{A}(\nu_{l}\to \nu_{l'}) =\sum_{i}U_{l'i}\int e^{i(\vec{p_{i}}'\vec{x}-
E'_{i}t)}
~f(\vec{p_{i}}'-\vec{p_{i}})~d^{3}p'~U^{*}_{li}
\end{equation}
Here $E'_{i}=\sqrt{(\vec{p_{i}}')^{2}+m^{2}_{i}}$ and the function
$f(\vec{p_{i}}'-\vec{p}_{i})$ has a sharp maximum at the point
$\vec{p'}_{i}=\vec{p}_{i}$. We assume that $\overline{|\vec{p_{i}}'-\vec{p}_{i}|}\ll p_{i}$.

Expanding $E'_{i}$ at the point $\vec{p_{i}}'=\vec{p}_{i}$ we have
\begin{equation}\label{wavepack1}
E'_{i}\simeq E_{i} +
(\vec{p_{i}}'-\vec{p_{i}})\cdot \vec{v_{i}},
\end{equation}
where $E_{i}=\sqrt{\vec{p_{i}}^{2}+m^{2}_{i}}$ and
\begin{equation}\label{wavepack2}
\vec{v}_{i}=\frac{\vec{p}_{i}}{E_{i}}
\end{equation}
Taking into account (\ref{wavepack1}) we find
\begin{equation}\label{wavepack3}
\int e^{i(\vec{p_{i}}'\vec{x}-
E'_{i}t)}
~f(\vec{p_{i}}'-\vec{p_{i}})~d^{3}p'= e^{-i(\vec{p_{i}}\vec{x}-
E_{i}t)}
~g(\vec{x}-\vec{v_{i}t}),
\end{equation}
where the amplitude $g(\vec{x}-\vec{v}_{i}t)$ is given by the expression
\begin{equation}\label{wavepack4}
g(\vec{x}-\vec{v}_{i}t)=\int e^{i\vec{q}~(\vec{x}-\vec{v}_{i}t)}~f(\vec{q})~d^{3}q.
\end{equation}
Notice that the wave packet  transition amplitude
differs from the amplitude in the plane wave approximation by the additional factor $g$. Because of the relativistic relation between momentum and energy this factor depends on the combination $\vec{x}-\vec{v}_{i}t$.

Usually it is assumed that the function $f(\vec{q})$   has the Gaussian form
\begin{equation}\label{wavepack5}
f(\vec{q})=N~e^{-\frac{q^{2}}{4\sigma_{p}^{2}}},
\end{equation}
where $\sigma_{p}$ is the width of the wave packet in the momentum space.
From (\ref{wavepack4}) and (\ref{wavepack5}) we find
\begin{equation}\label{wavepack6}
g(\vec{x}-\vec{v}_{i}t)=N (\frac{\pi}{\sigma_{x}^{2}})^{3/2}~e^
{-\frac{(\vec{x}-\vec{v}_{i}t)^{2}}{4\sigma_{x}^{2}}},
\end{equation}
where $\sigma_{x}=\frac{1}{2\sigma_{p}}$ characterizes the spacial width of the wave packet.

In the wave packet approach the probability of the transition $\nu_{l}\to \nu_{l'}$ is determined as a quantity obtained by integration over time (assuming that in neutrino oscillation experiments time is not measured)
\begin{equation}\label{wavepack7}
P(\nu_{l}\to \nu_{l'})=\int^{+\infty}_{-\infty}
|\mathcal{A}(\nu_{l}\to \nu_{l'})|^{2}dt
\end{equation}
 From (\ref{wavepack3}) and (\ref{wavepack7})  we find the following expression
for the normalized transition probability
\begin{equation}\label{wavepack8}
P(\nu_{l}\to \nu_{l'})=
\sum_{i,k}U_{l'i}U^{*}_{l'k}e^{i(p_{i}-p_{k})x}U^{*}_{li}U_{lk}
~e^{A_{ik}},
\end{equation}
where
\begin{equation}\label{wavepack9}
A_{ik}=-i(E_{i}-E_{k})x -\frac{1}{2\sigma_{x}^{2}}~
\left(\frac{\Delta m_{ik}^{2}}{4E^{2}}\right)^{2}x^{2}-\frac{1}{2}\sigma_{x}^{2}\xi^{2}
\left(\frac{\Delta m_{ik}^{2}}{2E}\right)^{2}.
\end{equation}
Here $E_{i}=E+\xi\frac{m^{2}_{t}}{2E}$ and $\xi$ is a constant of the order of one.

The factor $e^{A_{ik}}$ is the result of the integration over $t$. From the first term of the expression for $A_{ik}$ it is evident that
the Gaussian amplitude $g(\vec{x}-\vec{v}_{i}t)$ (after the integration over $t$) provides the equality  $t=x$. For usual neutrino oscillation experiments the second and the third terms of the expression for $A_{ik}$ are very small. In fact, let us introduce the coherence and oscillation lengths\footnote{We have $|v_{i}- v_{k}|L^{ik}_{\mathrm{coh}}\simeq \frac{|\Delta m^{2}_{ik}|}{2E^{2}}\sim \sigma_{x}$. Thus, the
coherence length characterizes such a distance between neutrino source and detector at which the distance between $\nu_{i}$ and
$\nu_{k}$ becomes comparable to the size of the wave packet.}
\begin{equation}\label{wavepack10}
L^{ik}_{\mathrm{coh}}=\frac{4\sqrt{2}\sigma_{x}E^{2}}{|\Delta m^{2}_{ik}|},\quad
L^{ik}_{\mathrm{osc}}=4\pi \frac{E}{|\Delta m^{2}_{ik}|}.
\end{equation}
The expression for the transition probability takes the form
\begin{equation}\label{wavepack11}
\mathcal{P}(\nu_{l}\to \nu_{l'})=\sum_{i,k}U_{l'i}U^{*}_{l'k}
e^{i\frac{\Delta m^{2}_{ik}}{2E}L}U^{*}_{li}U_{lk}~
e^{-(\frac{L}{L^{ik}_{\mathrm{coh}}})^{2}}~e^{-2\pi^{2}\xi^{2}
(\frac{\sigma_{x}}{L^{ik}_{\mathrm{osc}}})^{2}},
\end{equation}
where $x=L$ is the distance between neutrino source and neutrino detector.

We have
\begin{equation}\label{wavepack11}
L^{ik}_{\mathrm{coh}}=\frac{\sqrt{2}}{\pi}\sigma_{x}E L^{ik}_{\mathrm{osc}}.
\end{equation}
From this expression It follows that the coherence length is much larger than the oscillation length. Thus, for neutrino oscillation experiments with $L\simeq
L^{ik}_{\mathrm{osc}}$ the term $e^{-(\frac{L}{L^{ik}_{\mathrm{coh}}})^{2}}$ is practically equal to one.

Further we have\footnote{As we discussed before, because of this inequality coherent flavor neutrino states are produced.}
\begin{equation}\label{wavepack12}
  L^{ik}_{\mathrm{osc}}\gg  \sigma_{x}.
\end{equation}
Thus, the term $e^{-2\pi^{2}\xi^{2}
(\frac{\sigma_{x}}{L^{ik}_{\mathrm{osc}}})^{2}}$ is also practically equal to one.

We will finish this part with the following remarks:
\begin{itemize}
  \item Integration over time in the wave packet approach assumes that the time interval $t$ between neutrino production and detection is not measured in neutrino oscillation experiments. This is correct in the case of the atmospheric and reactor neutrino experiments because the time of neutrino production is not known in such experiments. However, in the case of the accelerator neutrino experiments (K2K, MINOS, T2K) neutrinos are produced in spills and the time of neutrino production is known. In these experiments  the time of neutrino production is measured and the time interval $t$ is known. For example, in the K2K experiment \cite{K2K} the measurement of  $ t=t_{SK}-t_{KEK}$, where $t_{SK}$ is the time of detection of neutrinos in the Super-Kamiokande detector and
$t_{KEK}$ is the time of the production of neutrinos at KEK, allowed to show that
  \begin{equation}\label{K2k}
   -0.2\leq |t-\frac{L}{c}|\leq 1.3~\mu \mathrm{s}.
\end{equation}
\item The wave packet approach assures the equality $t=L$ and the standard oscillation phase in the transition probability. Two additional exponential factors are very close to one for usual neutrino oscillation experiments. The effect of the decoherence  term could be important only for large cosmological distances.
\end{itemize}

In many papers (see \cite{Oscillations})  neutrinos, propagating  about 100 km (reactor $\nu$'s ) or  about 1000 km (atmospheric and accelerator $\nu$'s ), are considered as {\bf virtual neutrinos} in a Feynman diagram-like picture with the neutrino production process at one vertex and the neutrino absorption process in another vertex.
This approach gives the wave packet picture of neutrino oscillations with a transition probability which (before integration over $t$) depends on $x$ and $t$.

 In the standard $S$ matrix approach, which is based on the local quantum field theory, the transition amplitude is given by
\begin{equation}\label{transit}
    \langle f|S| i\rangle = \langle f|T(e^{-i\int \mathcal{H_{I}}(x)~d^{4}x})| i\rangle .
\end{equation}
where $\mathcal{H_{I}}(x)$ is the interaction Hamiltonian. Let us stress that
\begin{itemize}
  \item in all orders of the perturbation theory of the matrix element (\ref{transit}) integration over {\em the same (in our case weak) interaction region} is performed and  virtual particles belong to the same region. In the "virtual neutrino approach to neutrino oscillations" there are two interaction regions (production and detection) separated by a large macroscopic distance.
  \item in the standard $S$ matrix approach initial and final states are states of free particles considered at  {\em the same time} (correspondingly at $t\to -\infty$ and at $t\to +\infty$).  In the "virtual neutrino approach" initial and final states are states of particles at fixed space-time points separated by macroscopic distance and time.
\end{itemize}
This "virtual neutrino approach " can be considered as a model based on the combination of field theory and relativistic quantum mechanics. From our point of view, the applicability of this approach to neutrino oscillations requires experimental tests.

\section{Conclusions}
We have discussed different approaches to neutrino oscillations.
The QFT approach is based on the assumption that states of flavor neutrinos $\nu_{l}$ are mixed coherent states $|\nu_{l}\rangle=\sum_{i}U^{*}_{li}|\nu_{i}\rangle$. The evolution of flavor neutrino states in time is determined by the Schr\"odinger equation for quantum states. The QFT approach is based on the same general principles as the approach to $B^{0}\rightleftarrows \bar B^{0}$ etc. oscillations studied in detail at B-factories and other facilities. The important characteristic feature of this approach is the Mandelstam-Tamm time-energy uncertainty relation.

Other approaches are based on the assumption that in weak processes mixed coherent superpositions of plane waves or wave packets, describing neutrinos with different masses, are produced and detected. The evolution of mixed neutrino wave functions in space and time is determined by the Dirac equation.

Different approaches to neutrino oscillations lead to the same expression for the neutrino transition probability $\mathcal{P}(\nu_{l}\to \nu_{l'})$ in the standard neutrino oscillation experiments. In order to distinguish different approaches special neutrino oscillation experiments are necessary. Such experiments could be the recently discussed
M\"ossbauer neutrino experiments \cite{Raghavan,Potzel}.

As an example let us consider the recoilless M\"ossbauer
transition
\begin{equation}\label{mosstransitions}
 ^{3}\rm{H}\to ^{3}\rm{He}+\bar\nu_{e},\quad \bar\nu_{e}+
^{3}\rm{He}\to^{3}\rm{H}.
\end{equation}
in which a $\bar\nu_{e}$
with energy $\simeq$ 18.6 keV is produced and absorbed.

It was estimated in \cite{Raghavan}  that the uncertainty
of the energy of the  antineutrino in the M\"ossbauer
transition (\ref{mosstransitions}) is of the order
\begin{equation}\label{energyuncert}
(\Delta E)_{\mathrm{M}} \simeq 8.4\cdot 10^{-12}~\mathrm{eV}.
\end{equation}
Let us compare $(\Delta E)_{\mathrm{M}} $ given by (\ref{energyuncert}) with the quantity  $\frac{\Delta m_{A}^{2}}{2E}$ which could govern neutrino oscillations in (\ref{mosstransitions}). We have
\begin{equation}\label{energyuncert1}
\frac{\Delta m_{A}^{2}}{2E} \simeq 0.6\cdot 10^{-7}~\mathrm{eV}
\end{equation}
Thus, we have
\begin{equation}\label{energyuncert2}
(\Delta E)_{\mathrm{M}} \ll \frac{\Delta m_{A}^{2}}{2E}.
\end{equation}
This means that neutrino oscillations with the oscillation length given by $L^{\mathrm{A}}_{\mathrm{osc}}=4\pi \frac{E}{\Delta m_{A}^{2}}$ can not be observed
in the M\"ossbauer neutrino experiment  if the QFT approach is valid \cite{BFP}. This statement is in agreement with the time-energy uncertainty relation: the uncertainty of the energy in the M\"ossbauer transition is too small to fulfill the Mandelstam-Tamm time-energy uncertainty relation (\ref{uncertain3}) with $t\simeq L^{\mathrm{A}}_{\mathrm{osc}}$.

On the other side, if the space-time picture of neutrino oscillations is valid, neutrino oscillations with the oscillation length $L^{\mathrm{A}}_{\mathrm{osc}}$ will be observed
in the M\"ossbauer neutrino experiment \cite{Akhmedov}. In fact, for the oscillation phase we have in this case
\begin{equation}\label{energyuncert3}
(E_{3}-E_{2})-(p_{3}-p_{2})\simeq \frac{\Delta m_{A}^{2}}{2E}.
\end{equation}
In the space-time approach a significant change of the neutrino state at the distance
 $L^{\mathrm{A}}_{\mathrm{osc}}$ does not require a corresponding energy uncertainty. In other words, neutrino oscillations in the space-time approach do not necessarily follow the Mandelstam-Tamm uncertainty relation.

Neutrino oscillations (like $B^{0}\rightleftarrows \bar B^{0}$ etc. oscillations) are
an extremely important quantum phenomenon. Because of the interference nature of neutrino oscillations their
investigation allows to determine tiny neutrino mass-squared differences which are not reachable in other experiments.
The theory of neutrino oscillations is grounded on basic conceptions. The study of neutrino oscillations in a M\"ossbauer neutrino experiment with practically monoenergetic neutrinos would allow us to answer such fundamental questions of Quantum Theory as the problem of the existence of  mixed coherent flavor states, the problem of the evolution of the quantum states (in time or in space and time), the problem of the universal applicability of the time-energy uncertainty relation and others.

I would like to express my deep gratitude to Walter Potzel for numerous discussions. This work has been supported by funds of the DFG (Transregio 27: Neutrinos and Beyond), the Munich Cluster of Excellence (Origin and Structure of the Universe), and the Maier-Leibnitz-Laboratorium (Garching).

\end{document}